\documentclass[twocolumn,showpacs,preprintnumbers,prl]{revtex4}
\usepackage{graphicx}
\usepackage{dcolumn}
\usepackage{amssymb}
\usepackage{bm}

\begin{document}

\title{Geometry of Compressible and Incompressible Quantum Hall States: Application to Anisotropic Composite Fermion Liquids}
\author{Kun Yang}
\affiliation{National High Magnetic Field Laboratory and Department of Physics,
Florida State University, Tallahassee, Florida 32306, USA}

\begin{abstract}
Haldane's geometrical description of fractional quantum Hall states is generalized to compressible states. It is shown that anisotropy in the composite fermion Fermi surface is a direct reflection of this intrinsic geometry. A simple model is introduced in which the geometric parameter can be obtained exactly from other parameters including electron mass anisotropy. Our results compare favorably with recent measurements of anisotropy in composite fermion Fermi surface [D. Kamburov, Y. Liu, M. Shayegan, L. N. Pfeiffer, K. W. West, and K. W. Baldwin, Phys. Rev. Lett. {\bf 110}, 206801 (2013)]. Broader implications of our results are discussed.
\end{abstract}

\maketitle

Two-dimensional electron gas (2DEG) subject to a strong perpendicular magnetic field has been a constant source of surprises over the last thirty years, starting with the discovery of the fractional quantum
Hall (FQH) effect, and the elegant Laughlin wave function capturing its most basic physics. One of the important conceptual developments
is Haldane's recent observation\cite{Metric} that in contrast to common belief that the Laughlin {\em wave function} contains no variational parameter, it actually contains a hidden (continuous) geometrical degree of freedom characterizing the anisotropy in the correlation hole surrounding each electron. He further points out this anisotropy should be treated as a variational parameter in the presence of
either anisotropic electron band mass or anisotropic interaction. This family of Laughlin {\em states} (which includes the original Laughlin {\em wave function} as a special member) have been constructed explicitly and generalized to other FQH states\cite{qiu}, and the variational program has been carried out in some specific cases\cite{boyang,haowang,Apalkov}. These states may be relevant to the anisotropic
FQH state  observed at $\nu=7/3$\cite{xia}. Attempts\cite{duncanunpub,xiluo} have also been made to promote this geometric parameter to a {\em dynamical} degree of freedom, which captures the collective excitations of Laughlin and other FQH liquids.

While very exciting, the current state of affair is perhaps somewhat unsatisfactory in the following aspects. (i) As pointed out by Haldane\cite{Metric}, the geometric parameter $g$ (to be defined below) of a FQH state is determined by a compromise between the {\em independent} anisotropies of the effective mass tensor and interaction; thus far this has to be determined numerically and {\em approximately} by optimizing $g$ of a {\em trial} state (say Laughlin) with respect to a specific anisotropic Hamiltonian. It would be ideal to do this analytically and if possible, exactly. (ii) More importantly, it is unclear how to measure $g$ experimentally. In this work we address both of these points by showing the following. (a) We construct a simple model Hamiltonian in which the electrons interact with each other through a Gaussian potential. In this case $g$ can be determined exactly in closed form. With this result an approximate but analytic expression for $g$ is suggested for generic non-singular interactions. (b) This geometric degree of freedom also exists for {\em compressible} states. In particular, we show that for composite fermion (CF)\cite{jain} Fermi liquid states, $g$ characterizes the anisotropy of CF Fermi surface, which has been measured recently\cite{gokmen,kamburov,kamburov2}. Our results are in good qualitative agreement with existing measurements, and point to straightforward quantitative test of the theory in future experiments.

Consider the Hamiltonian
\begin{eqnarray}
H=T+V,
\end{eqnarray}
with the kinetic energy
\begin{eqnarray}
T=\sum_j{1\over 2m}\left[a(\Pi^j_x)^2+(\Pi^j_y)^2/a\right].
\label{eq:kinetic energy}
\end{eqnarray}
Here $j$ is electron index, $m/a$ and $ma$ are its effective mass along $x$ and $y$ directions respectively, while $m$ is their geometric mean. For isotropic effective mass we have $a=1$, and $a-1$ (assumed to be positive without loss of generality) is a measure of the mass anisotropy.
\begin{eqnarray}
{\bf \Pi}={\bf p}+{e\over c}{\bf A}({\bf r})
\end{eqnarray}
is the mechanical momentum, $\nabla\times {\bf A}({\bf r})=-B\hat{z}$ thus the electrons move in a uniform perpendicular magnetic field. The guiding center coordinates
\begin{eqnarray}
{\bf R}={\bf r}-(\ell^2/\hbar)\hat{z}\times{\bf \Pi}
\end{eqnarray}
commute with ${\bf \Pi}$. Here $\ell=\sqrt{\hbar c/(eB)}$ is the magnetic length. The interaction term
\begin{eqnarray}
V=\sum_{i < j} v({\bf r}_i-{\bf r}_j)=\sum_{i < j}\sum_{\bf q}v_{\bf q} e^{i{\bf q}\cdot({\bf r}_i-{\bf r}_j)},
\end{eqnarray}
where $v_{\bf q}$ is the Fourier transform of $v({\bf r})$.

In the large $B$ limit, Landau level (LL) spacing overwhelms $V$, and the electron motion is confined to a given LL. In this case it is appropriate to project $V$ onto a given LL that results in a reduced Hamiltonian involving the ${\bf R}$'s only:
\begin{eqnarray}
\tilde{V}=\sum_{i < j}\sum_{\bf q}v_{\bf q} e^{i{\bf q}\cdot({\bf R}_i-{\bf R}_j)}F_n({\bf q}, a),
\label{eq:Vprojected}
\end{eqnarray}
where
\begin{eqnarray}
&&F_n({\bf q}, a)=|\langle n|e^{i{\bf q}\cdot(\hat{z}\times{\bf \Pi})}|n\rangle|^2\\
&=&e^{-(aq_x^2+q_y^2/a)\ell^2/2}|L_n[(aq_x^2+q_y^2/a)\ell^2/2]|^2
\end{eqnarray}
is the form factor of the $n$th LL ($L_n$ is the $n$th Laguerre polynomial). We note this is the {\em only} place the effective mass anisotropy parameter $a$ enters $\tilde{V}$, which is what we need to solve. This is a highly non-trivial task due to the non-commutativity of different components of ${\bf R}$:
\begin{eqnarray}
[R_x, R_y]=-i\ell^2.
\end{eqnarray}
For simplicity and later comparison with experiments of Ref. \cite{gokmen,kamburov,kamburov2}, we will focus on a partially filled lowest LL (LLL) where $n=0$, so
\begin{eqnarray}
F_0({\bf q}, a)=e^{-(aq_x^2+q_y^2/a)\ell^2/2}.
\end{eqnarray}
We will also work mostly with an {\em isotropic} interaction such that $v({\bf r})=v(r)$ and $v_{\bf q}=v_q$.

The key ingredient that allows for further progress below is the choice of a Gaussian potential
\begin{eqnarray}
v(r)&=&v(0)e^{-r^2/(2s^2)},\\
v_q&=&v_0e^{-q^2s^2/2},
\end{eqnarray}
where $s$ is the range of the potential in real space. With this choice the LLL-projected Hamiltonian $\tilde{V}$ of Eq. (\ref{eq:Vprojected}) becomes
\begin{eqnarray}
\tilde{V}_g=\sum_{i < j}\sum_{\bf q}v_0 e^{-\tilde{q}^2\tilde{s}^2/2} e^{i{\bf q}\cdot({\bf R}_i-{\bf R}_j)},
\label{eq:Vg}
\end{eqnarray}
where $\tilde{s}=[(a\ell^2+s^2)(\ell^2/a+s^2)]^{1/4}$ is a new length scale, while
\begin{eqnarray}
\tilde{q}^2=g q_x^2 + q_y^2/g,
\end{eqnarray}
in which the crucial geometric parameter
\begin{eqnarray}
g=\sqrt{(a\ell^2+s^2)/(\ell^2/a+s^2)}.
\label{eq:g}
\end{eqnarray}
We note $g\rightarrow a$ for $s\ll \ell$ while $g\rightarrow 1$ for $s\gg \ell$, and in general we have
\begin{eqnarray}
1 < g < a.
\label{inequality}
\end{eqnarray}
Eq. (\ref{eq:g}) is the central result of this work.

To gain insight into the solution of the LLL-projected Hamiltonian (\ref{eq:Vg}), we note that $\tilde{V}_g$ can be obtained from its isotropic version ($g=1$) by performing a unitary transformation that is a member of the area-preserving diffeomorphism\cite{halvisc,RR}:
\begin{eqnarray}
\tilde{V}_g=O^\dagger[\lambda(g)]\tilde{V}_{g=1} O[\lambda(g)],
\label{eq:Utransform}
\end{eqnarray}
where
\begin{eqnarray}
O(\lambda)=e^{{i\lambda\over 2\ell^2}\sum_j R^j_x R^j_y},
\end{eqnarray}
and
\begin{eqnarray}
\lambda(g)=-{1\over 2}\ln g.
\end{eqnarray}
This is a consequence of the properties that
\begin{eqnarray}
O^\dagger(\lambda)R_x^i O(\lambda)&=&e^{\lambda} R_x^i;\\
O^\dagger(\lambda)R_y^i O(\lambda)&=&e^{-\lambda} R_y^i.
\end{eqnarray}
We can thus obtain the solution of $\tilde{V}_g$ by performing a proper unitary transformation on the solution of $\tilde{V}_1$, which is an {\em isotropic} LLL projected Hamiltonian. In the following we discuss some special cases.

First consider Laughlin filling factors $\nu={1\over 2p+1}$. In this case because our Gaussian interaction is purely repulsive and a monotonically decreasing function of $r$, based on the extreme robustness of the Laughlin state for such interactions established by decades of numerical studies we expect the ground state of the system to be a FQH state very accurately approximated by the original isotropic Laughlin state $|\Psi^L\rangle \approx |\Psi_{g=1}\rangle $, as long as $s$ is not much larger than $\ell$. As a consequence of (\ref{eq:Utransform}) we have the ground state of the anisotropic case to be
\begin{eqnarray}
|\Psi_{g}\rangle=O^\dagger[\lambda(g)]|\Psi_{g=1}\rangle \approx |\Psi^L_g\rangle,
\label{eq:transform}
\end{eqnarray}
where
\begin{eqnarray}
|\Psi^L_{g}\rangle=O^\dagger[\lambda(g)]|\Psi^L\rangle
\end{eqnarray}
is an {\em anisotropic} Laughlin state\cite{Metric,qiu}.

Now consider even denominator filling factors $\nu={1\over 2p}$, where the electrons are expected to form a composite fermion Fermi liquid-like state\cite{hlr}. In this case the ground state is well approximated in the isotropic case by the following wavefunction\cite{rr}
\begin{eqnarray}
|\Psi^{FL}_{g=1}\rangle = \det(M_{jl}) |\Psi^{BL}\rangle,
\end{eqnarray}
where $|\Psi^{BL}\rangle$ is the isotropic {\em bosonic} Laughlin state at $\nu={1\over 2p}$, $\det(M_{jl})$ is the determinant of a matrix whose entries are
\begin{eqnarray}
M_{jl}=e^{i{\bf k}_j\cdot{\bf R}_l},
\end{eqnarray}
and the set $\{{\bf k}_j\}$ form a circular or isotropic Fermi sea.

Similar to Eq. (\ref{eq:transform}), for the anisotropic case $g\ne 1$ the ground state is well approximated by an anisotropic CF Fermi sea state:
\begin{eqnarray}
&&|\Psi^{FL}_{g}\rangle = O^\dagger[\lambda(g)]|\Psi^{FL}_{g=1}\rangle \\
&=& \left\{O^\dagger[\lambda(g)]\det(M_{jl})O[\lambda(g)]\right\} O^\dagger[\lambda(g)]|\Psi^{BL}\rangle\\
&=&\det(M'_{jl})|\Psi_g^{BL}\rangle,
\end{eqnarray}
where $|\Psi_g^{BL}\rangle$ is an anisotropic bosonic Laughlin state, and
\begin{eqnarray}
M'_{jl}=O^\dagger[\lambda(g)]M_{jl}O[\lambda(g)]=e^{i{\bf k}'_j\cdot{\bf R}_l},
\end{eqnarray}
with
\begin{eqnarray}
{\bf k}'_j={k_j^x \hat{x}\over\sqrt{g}}+ \sqrt{g}k_j^y\hat{y}.
\end{eqnarray}

We see the anisotropic CF Fermi liquid state $|\Psi^{FL}_{g}\rangle$ differs from its isotropic version in two aspects. (i) the Laughlin factor needs to be replaced by its anisotropic version. Since this factor describes flux attachment\cite{hlr} and introduces a correlation hole around each CF, this implies a distortion of this correlation hole, similar to what happens in the anisotropic Laughlin states\cite{qiu}. (ii) More importantly, the  Slater determinant of the (LLL projected) plane wave factors is now formed by a set of plane waves with wave vectors $\{{\bf k}'_j\}$, which form an anisotropic Fermi sea of elliptic shape, with the ratio between long and short axes
\begin{eqnarray}
k^{CF}_{fy}/k^{CF}_{fx}=g.
\label{eq:cf anisotropy}
\end{eqnarray}
This is {\em smaller} than the corresponding anisotropy at zero magnetic field:
\begin{eqnarray}
k_{fy}/k_{fx}=a
\label{eq:electron anisotropy}
\end{eqnarray}
due to (\ref{inequality}).

A few comments are no in order. (i) The analyzes above can be generalized straightforwardly for anisotropic interactions of the Gaussian type:
\begin{eqnarray}
v_{\bf q}=v_0e^{-q_{\alpha}q_{\beta}s^2_{\alpha\beta}/2},
\end{eqnarray}
where $s^2_{\alpha\beta}$ is a symmetric $2\times 2$ tensor, and repeated indices are summed over in the above.
In this case because the anisotropy in the effective mass and interaction have independent orientations, they both need to be characterized by a real anisotropic parameter as well as an orientation angle, which can be combined into a single complex parameter. The same is true for the geometric parameter $g$ of the resultant state. (ii) For a generic form of electron-electron interaction, an {\em exact} relation between the geometric parameter $g$ and effective mass anisotropy $a$ is not available. However for generic non-singular interactions, it is possible to extract a length scale $s$ by inspecting the short-distance behavior of $v(r)$:
\begin{eqnarray}
v(r)=v(0)[1-r^2/(2s^2)+o(r^2)].
\end{eqnarray}
A natural {\em approximation} for $g$ is Eq. (\ref{eq:g}) with $s$ defined above. We note for the $1/r$ Coulomb interaction such a length scale does not exist and Eq. (\ref{eq:g}) cannot be used; numerical study is thus needed to extract the geometric parameter $g$. For $\nu=1/3$ it was found\cite{boyang,note} that $g^2\approx 1+ 0.4(a^2-1)$ thus the inequality (\ref{inequality}) is satisfied. We note in reality the Coulomb interaction is regularized at short distance by a finite width of the quantum well and thus a length scale $s$ does exist.

We now compare our results with recent measurements\cite{gokmen,kamburov,kamburov2} of CF Fermi surface anisotropy that results from electron effective mass anisotropy, which, as we have shown, directly probe the geometry of the corresponding CF liquid state. It was found that anisotropy in electron dispersion and corresponding Fermi surface (in the absence of perpendicular magnetic field) indeed induce an anisotropy in the CF Fermi surface, but the anisotropy is {\em always} smaller for the latter\cite{gokmen,kamburov}. This is clearly inconsistent with an earlier theory\cite{balagruv} but in agreement with our inequality (\ref{inequality}). More quantitatively, it is found that the CF anisotropy is often {\em much weaker} than the electrons; for example in one case\cite{kamburov} it was found $a\approx 3$ while $g \approx 1.2$. According to Eq. (\ref{eq:g}) this implies $s\approx 2.4 \ell$, which is not unreasonable if we identify $s$ as the well width of 175 \AA, coupled with $\ell=67$ \AA ~ at $B=B_\bot =14$T. Of course, the equation (\ref{eq:g}) can be tested systematically by measuring the dependence of $g$ on both the electron effective mass anisotropy $a$, and (perpendicular component of) magnetic field (through the magnetic length $\ell$). A note of caution is in Refs. \cite{kamburov,kamburov2} $a$ is induced by an in-plane magnetic field, whose effects can be somewhat complicated, and the effective mass approximation may break down when the in-plane magnetic field get too strong. Measuring the dependence of $g$ on perpendicular magnetic field requires samples with tunable density so as to stay at a a fixed filling factor (say 1/2).

In summary, we broaden the scope of the geometric description of FQH effect significantly, by including compressible electron liquids in the FQH regime in this description. More importantly, our work reveals how to probe this geometry {\em experimentally}, at least in the presence of a composite fermion Fermi surface. In a more general context, we note strongly correlated states with anisotropic (or nematic) Fermi surfaces are of strong current interest in other correlated electron systems (in particular high-$T_c$ cuprates)\cite{fradkin}. We have introduced and thoroughly investigated a simple model which clearly reveals how such anisotropic Fermi surface results from the interplay between single particle band structure and electron-electron interaction, which should be of broad interest.

This work was supported by DOE grant No. DE-SC0002140.


\begin{references}


\bibitem{Metric} F. D. M. Haldane, Phys. Rev. Lett. {\bf 107}, 116801 (2011).

\bibitem{qiu} R.-Z. Qiu, F. D. M. Haldane, Xin Wan, Kun Yang, Su Yi, Phys. Rev. B {\bf 85}, 115308 (2012).

\bibitem{boyang} Bo Yang, Z. Papic, E. H. Rezayi, R. N. Bhatt, and F. D. M. Haldane, Phys. Rev. B {\bf 85}, 165318 (2012).

\bibitem{haowang} Hao Wang, Rajesh Narayanan, Xin Wan, and Fuchun Zhang, Phys. Rev. B {\bf 86}, 035122 (2012).

\bibitem{Apalkov} V. M. Apalkov and Tapash Chakraborty, arXiv:1306.2408.

\bibitem{xia} J. Xia, J. P. Eisenstein, L. N. Pfeiffer, K. W. West,
Nature Phys. {\bf 7}, 845 (2011).

\bibitem{duncanunpub} F. D. M. Haldane, unpublished.

\bibitem{xiluo} Xi Luo, Yong-Shi Wu, and Yue Yu, arXiv:1308.2481.

\bibitem{jain} J. K. Jain, {\em Composite Fermions}, Cambridge University Press (2007).

\bibitem{gokmen} T. Gokmen, Medini Padmanabhan, M. Shayegan,  Nature Phys. {\bf 6}, 621 (2010).

\bibitem{kamburov} D. Kamburov, Yang Liu, M. Shayegan, L. N. Pfeiffer, K. W. West, and K. W. Baldwin, Phys. Rev. Lett. {\bf 110}, 206801 (2013).

\bibitem{kamburov2} D. Kamburov, M. A. Mueed, M. Shayegan, L. N. Pfeiffer, K. W. West, K. W. Baldwin, J. J. D. Lee, R. Winkler, arXiv:1306.3537; arXiv:1308.1728.

\bibitem{halvisc} F. D. M. Haldane, arXiv:0906.1854 (unpublished).

\bibitem{RR} N. Read and E. H. Rezayi, Phys. Rev. B {\bf 84}, 085316 (2011).

\bibitem{hlr} B. I. Halperin, P. A. Lee and N. Read, Phys. Rev. B {\bf 47}, 7312 (1993).

\bibitem{rr} E. H. Rezayi and N. Read, Phys. Rev. Lett. {\bf 72}, 900 (1994); E. H. Rezayi and F. D. M. Haldane, Phys. Rev. Lett. {\bf 84}, 4685 (2000).

\bibitem{note} See Fig. 3 of Ref. \onlinecite{boyang}. Their $\alpha$ and $\alpha_m$ correspond to our $g^2$ and $a^2$.

\bibitem{balagruv} D. B. Balagurov and Yu. E. Lozovik, Phys. Rev. B {\bf 62}, 1481 (2000).

\bibitem{fradkin} E. Fradkin, S.A. Kivelson, M.J. Lawler, J.P. Eisenstein, and A.P. Mackenzie, Annual Review of Condensed Matter Physics {\bf 1}, 153 (2010).


\end{references}
\end{document}